\documentclass[journal]{IEEEtran}
\usepackage{color,array}
\usepackage[english]{babel}
\usepackage{inputenc}
\usepackage{algorithm}
\usepackage{algorithmicx}

\usepackage{algcompatible}
\usepackage{algpseudocode}
\usepackage{makecell}
\usepackage{cite}
\usepackage{pbox}
\usepackage{amsmath,amssymb}

\usepackage[pdftex]{graphicx}
\usepackage[figurename=Fig.,font=small,skip=2 pt]{caption}
\graphicspath{{../pdf/}{../jpeg/}}
\graphicspath{ {./img/}}
\graphicspath{ {./img1/}}
\DeclareGraphicsExtensions{.pdf,.jpeg,.png}
\usepackage{epstopdf}
\usepackage{multirow}
\usepackage{accents}
\newcommand{\justified}{%
  \rightskip\z@skip%
  \leftskip\z@skip}
\usepackage{multicol}
\usepackage{enumerate}
\usepackage{array}
\usepackage{color,soul}
\usepackage{pbox}

\usepackage{comment}
\usepackage{url}
\usepackage{relsize}

\usepackage{stfloats}
\usepackage{amssymb}

\usepackage{setspace}
\usepackage{amsmath}
\usepackage{float}
\usepackage[table]{xcolor}
\definecolor{Gray}{gray}{0.85}
\usepackage{verbatim} 
\usepackage{color, colortbl}
\definecolor{Gray}{gray}{0.92}

\def\l{\left}   
\def\r{\right}   

\include{macros}

\begin{document}

\title{Knowledge Transfer-based Evolutionary Deep Neural Network for Intelligent Fault Diagnosis}
\author{Arun Kumar Sharma,~\IEEEmembership{Student Member,~IEEE and}
        Nishchal Kumar Verma,~\IEEEmembership{Senior Member,~IEEE}
      
\thanks{Arun Kumar Sharma and Nishchal Kumar Verma are with the Dept. of Electrical Engineering, Indian Institute of Technology, Kanpur, India.
e-mail: arnksh@iitk.ac.in and nishchal@iitk.ac.in}
}

\maketitle

\begin{abstract}
A faster response with commendable accuracy in intelligent systems is essential for the reliability and smooth operations of industrial machines. Two main challenges affect the design of such intelligent systems: (i) the selection of a suitable model and (ii) domain adaptation if there is a continuous change in operating conditions. Therefore, we propose an evolutionary Net2Net transformation (EvoN2N) that finds the best suitable DNN architecture with limited availability of labeled data samples. Net2Net transformation-based quick learning algorithm has been used in the evolutionary framework of Non-dominated sorting genetic algorithm II to obtain the best DNN architecture. Net2Net transformation-based quick learning algorithm uses the concept of knowledge transfer from one generation to the next for faster fitness evaluation. The proposed framework can obtain the best model for intelligent fault diagnosis without a long and time-consuming search process. The proposed framework has been validated on the Case Western Reserve University dataset, the Paderborn University dataset, and the gearbox fault detection dataset under different operating conditions. The best models obtained are capable of demonstrating an excellent diagnostic performance and classification accuracy of almost up to 100\% for most of the operating conditions.
\end{abstract}
\IEEEpeerreviewmaketitle


\section{Introduction}
\label{intro}
\IEEEPARstart {W}{ith} the increase in complexity in modern industrial machinery, continuous monitoring has become a major concern to ensure the reliability, smooth operation, and productivity of systems. Therefore, continuous fault diagnosis, also called condition-based monitoring (CBM), is an essential requirement of today's industries and has gained much attention from researchers worldwide \cite{10504763, HU2025102963, 10143698}. Progress in the area of fault diagnosis of industrial machines can be divided roughly into three stages: (i) traditional methods, (ii) fault diagnosis based on signal processing and analysis, and (iii) intelligent fault diagnosis. 

Traditional fault diagnosis methods mainly rely on experience with machine running conditions. The signal processing and analysis-based fault diagnosis requires complex mathematical analysis to identify specific changes in the machine signals like current, vibration, temperature, acoustic transmissions, etc. \cite{sr2, sr11}. Therefore, these methods are not suitable for continuous monitoring in modern industrial scenarios. 
In recent decades, intelligent fault diagnosis has been the most investigated method due to the learning capability of machine learning models \cite{rn1, svm, sparseAE, rzhao, newfd1, newfd5}. Support vector machine (SVM) \cite{svm} has been reported with outstanding performance for intelligent pattern recognition. However, it fails to perform well for big data classification with high sparsity. For such problems, applications of deep neural networks (DNN) have been reported to be very effective \cite{sparseAE, rzhao}. DNN is highly capable of hierarchical feature transformation of raw data to linearly separable features. Recently, several feature extraction methods using genetic algorithms have been reported, \cite{EC_feature, featureConst, AutFeatureExtr} for the training of DNN. However, for continuous monitoring under changeable machine working conditions, training a new DNN for the new dataset is an uneconomical and time-consuming task. 

The concept of transfer learning accelerated the DNN training process for the new data domain by initializing the weight parameters using knowledge of the pre-trained model on the source data \cite{pratt, sjPan}. Further, to solve the problem of domain shift, domain adaptation based on transfer component analysis was introduced in 2011 by S.J. Pan \cite{sjPan2011}. Later, the concept of domain adaptation using minimization of Maximum Mean Discrepancy (MMD) gained much attention by various researchers \cite{longRTRL, lwen, cross-domain, aks_quick, dagcn}. Long \textit{et. al.} \cite{longRTRL} used the principle of structural risk minimization and regularization theory to obtain a mechanism of adaptation regularization for cross-domain learning. L. Wen \textit{et. al.} \cite{lwen} suggested a fine-tuning method by minimizing the classification loss and the MMD term calculated for labeled source data and the labeled target data. The performance of all these methods is greatly affected by the selection of model architecture. Therefore, fault diagnosis under variable operating conditions of machines demands a fast network architecture search method capable of finding the best suitable model for the given working condition using a small amount of leveled target data.

The motivation behind this work is to develop an automatic architecture optimization algorithm for DNN to classify accurately in the new data domain. Various methods for the neural architecture search have been reported in literature \cite{NIPS2011, BJ,  BB_nn_rein, c1rv1, sun_pso, Ysun2020, EvoDCNN, newfd1, PHAN2024101573}. \cite{NIPS2011,  BJ} suggested the use of random search and greedy sequential algorithm for the optimization of hyperparameters of neural networks and deep belief networks. Random search or sequential search algorithms explore all possible architectures in a given search space.  Therefore, exploring for the best suitable architecture takes too long.  Baker \textit{et. al.} \cite{BB_nn_rein} introduced a guided search algorithm using a Q-learning agent for the automatic selection of CNN architecture.   In recent decades, genetic algorithms (GA) have gained much attention for hyperparameter optimization due to their multi-constraint optimization capabilities \cite{EvoDCNN, sun_pso, Ysun2020}. However, the application of GA to deep learning algorithms is a complex problem due to a large number of constraint variables and a time-consuming fitness evaluation (training and testing) process. Further, Luo \textit{et. al.} \cite{newfd1} proposed Wasserstein GAN meta-learning (WGANML) that uses the concept of the game between the generative and discriminative models to generate missing samples, followed by meta-learning for parameter adjustment for better results. Sun \textit{et. al.} \cite{Ysun2020} designed a variable-length gene encoding technique to encode the CNN architecture and adopted the fitness evaluation strategy of training a CNN model for a few iterations to find the best of the generation. Each individual (CNN model) in the population is initialized using Gaussian distribution and trained from scratch. This method requires the training of CNN models from scratch at each generation. Therefore, the evolution of architecture for the best performance becomes a time-consuming process. 

Since most of the architecture search methods rely on training and evaluating DNN with different architectures and concluding with the best model, the major challenge in the architecture search process is the quick fitness evaluation strategy. The concept of function-preserving-based network transformation gives a breakthrough for transforming the existing model to a new model \cite{aks_quick, Ian, tl-gdbn}. The network transformation method proposed in \cite{aks_quick} paves a way to quickly train a new model for the new target dataset through a network-to-network knowledge transformation. Based on \cite{aks_quick}, we propose an evolutionary DNN architecture with a quick fitness evaluation framework using knowledge transfer from generation to generation. The key contributions are summarized below:
\begin{enumerate}[i)]
    \item An evolutionary deep neural network with network-to-network transformation (EvoN2N) for automatic architecture optimization is proposed. The fitness evaluation framework is based on knowledge transfer with domain adaptation from one network model to another. An initial (source) model trained on source data is used to initialize a new model followed by fine-tuning using a small number of samples of the target dataset.
    \item The proposed framework includes a two-step crossover technique for two chromosomes with different lengths. This mechanism helps the algorithm to exploit the search space of depth and width of the DNN architecture, simultaneously.
\end{enumerate}

The organization of the rest of the paper is as follows. Section \ref{prob_formulation} defines the problem objective of automatic architecture search. Section \ref{theoretical} briefly introduces the theoretical background of evolutionary algorithms, multi-objective optimization, DNN, and knowledge transfer from a DNN to another through the function-preserving principle. Section \ref{proposed} describes the proposed EvoN2N. Section \ref{experi} discusses the effectiveness of the proposed framework and its comparison with state-of-the-art methods on (i) Case Western Reserve University (CWRU) dataset \cite{cwru}, (ii) the Paderborn University (PBU) dataset \cite{paderborn} and (iii) the gearbox fault detection dataset \cite{gfd} under different operating conditions. Finally, Section \ref{conclusion} concludes the work.

\section{Theoretical Background}\label{theoretical}
\subsection{Multi-objective Genetic Algorithm}\label{sec:nsga}
Multi-Objective Genetic Algorithm (MOGA) is a heuristic search-based optimization technique under multi-constraint to find multiple optimal solutions called Pareto-optimal solution. A set of multiple solutions is obtained, out of which, any one solution cannot be said to be better without some additional information. For such problems, several MOGAs have been reported in various publications \cite{moea2, nsga, nsga-iii, 10807400}. The non-dominated sorting genetic algorithm (NSGA-II) \cite{nsga} has gained much attention due to its fast sorting methodology. Our problem consists of unconstrained optimization of the depth \& width of a DNN architecture for maximum diagnostic performance while keeping the number of total trainable parameters minimal. Therefore, we have considered architecture optimization as a multi-objective optimization problem instead of a single-objective optimization problem. We have selected the NSGA-II framework for the evolution of DNN architecture, considering the objectives (i) to maximize the validation accuracy and (ii) to minimize the total number of weight matrices. 

\subsection{Domain Adaptation}\label{sec:dom_ada}
\begin{figure}[!ht]
\centering
\includegraphics[width=8cm]{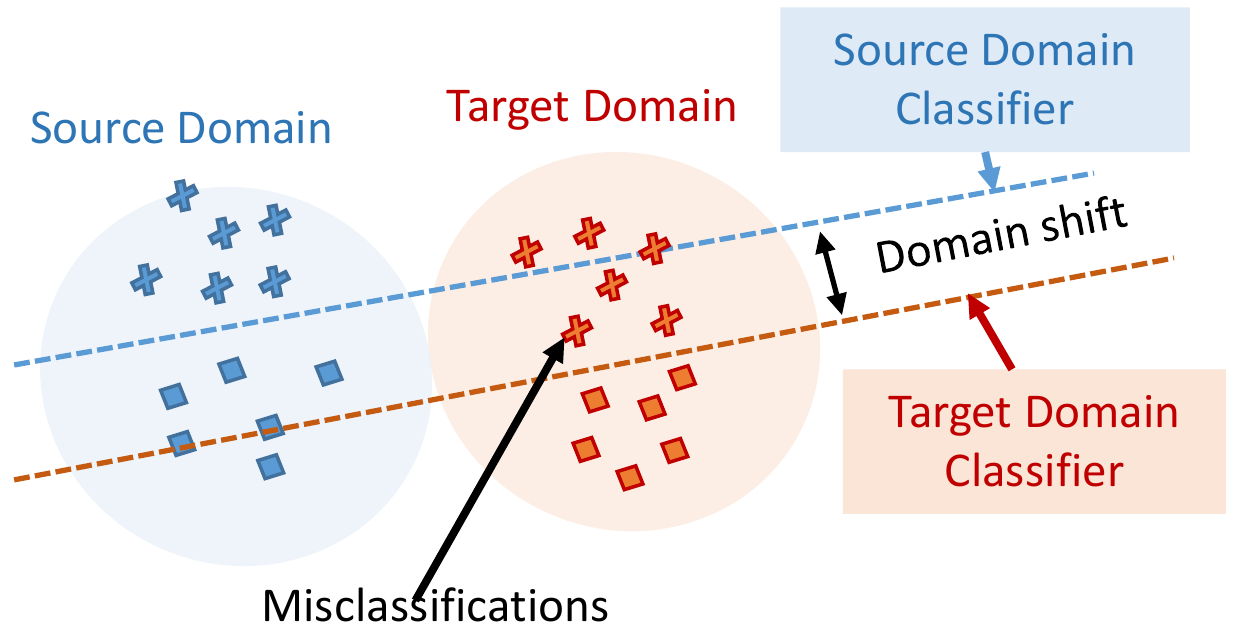}
\caption{Domain shift between source and target dataset: A slight change in the data distribution from the source domain (blue) to the target domain (red) causes miss-classification and needs fine-tuning of the model for domain adjustment.}
\label{fig:doAdapt}
\end{figure}

If the target datasets have different distributions from the source data, the DNN trained on the source data fails to classify correctly on the target data. The diagnostic model trained on the source data needs to be fine-tuned using the target data to shift the classifier shown in Fig. \ref{fig:doAdapt}.

Maximum Mean Discrepancy (MMD) minimization is the most popular method for domain adaptation to the change in distribution. It measures the non-parametric distance for the domain shift on the reproducing kernel Hilbert space (RKHS). \cite{rkhs}.

\section{Proposed Methodology}
\subsection{Problem Statement}\label{prob_formulation} A DNN is a multi-layered neural network formed by stacked auto-encoder (SAE) \cite{bengio} with softmax classifier as output layer. DNN has the capability of highly non-linear function approximation. Each layer of SAE is trained by a greedy layer unsupervised learning mechanism followed by stacking together to form SAE. Now, SAE with softmax classifier at the output layer is fine-tuned by gradient descent using a labeled dataset.
Let $W^{s}$ be the teacher (source) model trained on a dataset $\mathcal{D}^{s}=\l(X^{s}, Y^{s}\r)$ from the source domain and $\mathcal{D}^{tr}=\l(X^{tr}, Y^{tr}\r)$, $\mathcal{D}^{val}=\l(X^{val}, Y^{val}\r)$, \& $\mathcal{D}^{te}=\l(X^{te}, Y^{te}\r)$ are the training dataset, the validation dataset, \& the test dataset respectively from the target dataset $\mathcal{D}^{t}=\l(X^{t}, Y^{t}\r)$. The objective of fault diagnosis via automatic DNN architecture search can mathematically be expressed as
\begin{flalign}
   & P \; = \; \mathcal{G}(W^{s}) \\
   & W^{t}_{best} =  \mathcal{H}\l(P,\; \mathcal{D}^{tr}, \;\mathcal{D}^{val}\r) \\
  &  \hat{Y}^{te} \; = \; \mathcal{F}\l(W^{t}_{best},\; X^{te}\r)
\end{flalign}
where $\mathcal{G}(.)$ generates a set ($P$) of DNN model with different architecture by transforming the model $W^{s}$, $\mathcal{H}(.)$ is the function to search and find the best model $W^{t}_{best}$ with optimal parameters and $\mathcal{F}(.)$ is the feed-forward sequential function of the DNN to predict the fault class $\hat{Y}^{te}$ for the given test data $X^{te}$.

\subsection{Proposed Framework}\label{proposed} This section presents the proposed methodology of the evolutionary Net2Net transformation (EvoN2N) in detail. Let $\Psi^s,\; \Psi^t, \; {\cal D}^s = (X^s, Y^s),$ and  ${\cal D}^t = (X^t, Y^t)$ be the initial (source) DNN model, the target DNN model, source dataset, and the target dataset respectively, the main framework of the NSGA-II based evolution and training of the DNN architecture is formulated as in Algorithm \ref{algo:proposed}. Detailed discussions on the implementation procedure for each step are presented in the following sections.
\begin{figure*}[!ht]
\centering
\includegraphics[width=\textwidth]{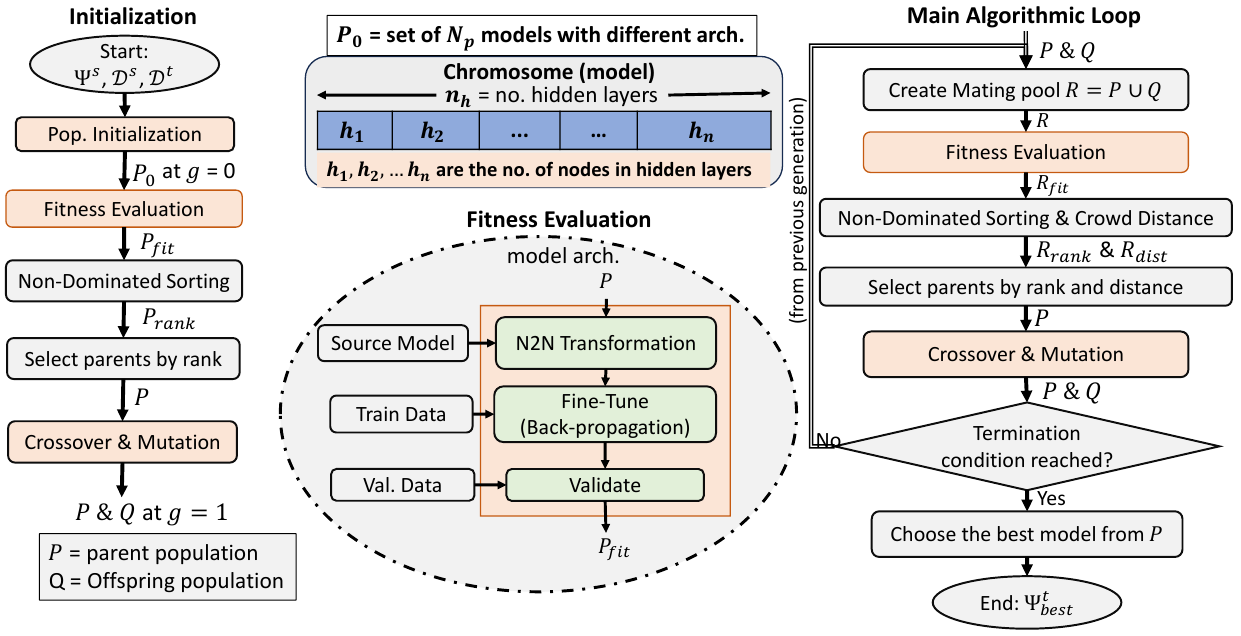}
\caption{Building blocks of NSGA-II frameworks for neural network architecture optimization using N2N transformation-based quick learning mechanism.}
\label{Fig: nsga_flow}
\end{figure*}

\begin{algorithm*}[!ht]
\caption{Main Framework of the EvoN2N}
\label{algo:proposed}
\begin{algorithmic}[1]
\item $Inputs \longleftarrow (\Psi^s,\;\mathcal{D}^{s},\; \mathcal{D}^{t})$ \hspace{0.5cm} //(initial model, source dataset, target dataset).
\item $Gen \longleftarrow 0$ \hspace{3.2cm}//Set generation count = 0;
\item $P_0\longleftarrow \textbf{InitializePopulation}(N_p\;[1,\;n_{max}],\; [h_{max}, h_{min}])$ \hspace{0.5cm} //Use \textbf{Algorithm. \ref{algo:popln}}.
\item $P_{fit},\, \Psi^t_{best}\longleftarrow \textbf{EvalFitness}(P_0, \Psi^s, \mathcal{D}^{s}, \mathcal{D}^{t}) $  //Evaluate fitness using \textbf{Algorithm \ref{algo:evafit}}.
\item $P_{rank}\longleftarrow NonDominatedSorting(P_{fit})$ \hspace{0.05cm} //Ranking with non-dominated sorting \cite{nsga}.
\item $P \longleftarrow \textbf{SelectParents}(P_0, \; P_{rank})\,$ \hspace{0.5cm}//\textbf{Binary tournament selection (Algo. \ref{algo:tour})}.
\item $Q \longleftarrow \textbf{CrossOverMutation}(P)$ \hspace{0.1cm} //Apply crossover and mutation (\textbf{Section \ref{sec:cross_mut}}).
\item $\Psi^s \longleftarrow \Psi^t_{best}$ \hspace{0.2cm}//Set the current best model as initial model for the next generation. 
\While{$Gen \leq MaxGeneration$}
\item $R  \longleftarrow (P \cup Q)$ \hspace{0.4cm}  //Combine the parent population ($P$) \& the child population ($Q$).
\item $R_{fit},\, \Psi^t_{best}\longleftarrow \textbf{EvalFitness}(R, \Psi^s, \mathcal{D}^{s}, \mathcal{D}^{t}) $ \hspace{0.1cm}  //Evaluate fitness using \textbf{Algorithm \ref{algo:evafit}}.
\item $R_{rank}\longleftarrow NonDominatedSorting(R_{fit})$ \hspace{0.05cm} //Ranking with non-dominated sorting \cite{nsga}
\item $R_{crowd}\longleftarrow CrowdingDistances(R, \,R_{rank},\,R_{fit})$ \hspace{5pt}  //Crowding distances of individuals in population set $R$ \cite{nsga}.
\item $P \longleftarrow \textbf{SelectParents}(R, R_{crowd}, R_{rank})$ //Select parents by crowding distance and rank.
\item $Q \longleftarrow \textbf{CrossOverMutation}(P)$  \hspace{0.1cm} //Apply crossover and mutation (\textbf{Section \ref{sec:cross_mut}}).
\item $\Psi^s \longleftarrow \Psi^t_{best}$ \hspace{0.2cm}//Set the current best model as initial model for the next generation. 
\item $Gen \longleftarrow Gen +1$ \hspace{0.9cm} //Update the generation counter
\EndWhile
\item Return: $\textrm{\textbf{Best Model}}\longleftarrow \Psi^t_{best}$ \hspace{1cm} //Best model of last generation.
\end{algorithmic}
\end{algorithm*}

\subsubsection{Gene Encoding and Population Initialization}\label{sec:genEncod} For optimal architecture search, each chromosome should be of variable length and contain the information of (i) depth: number of hidden layers ($n_h$) and (ii) width: number of nodes in each layer ($h_1, h_2, h_3, ...$).  The gene encoding with different lengths has been illustrated in Fig. \ref{fig:cross}. The length of the chromosome represents the depth of the network and the value of each gene represents the number of hidden nodes in each layer. Since the size of the chromosomes is variable, information about the network architecture is encoded using the real-coded approach. The procedure for population initialization is provided in steps 3-8 of Algorithm \ref{algo:proposed}. The population size, the maximum range for the depth of the network, and the maximum range for the number of nodes in a hidden layer are $N_p, \;{n_h}\in[1,\;n_{max}],\; \&\; h\in [h_{max},\; h_{min}]$ respectively.
\begin{algorithm}[!ht]
\caption{Initialize Population for EvoN2N}\label{algo:popln}
\begin{algorithmic}[1]
\item $Inputs \longleftarrow \l(N, \;[n_{min},\;n_{max}],\; [h_{max}, h_{min}]\r)$
\item $H\longleftarrow$ Generate $N$ random integers between $[1\rightarrow N_h]$.
\For{p = 1:N}
\item $h\longleftarrow H(p)$ : depth of $p^{th}$ chromosome
\item $P\{p\} \longleftarrow $ generate $h$ random integers $\in [h_{min}, h_{max}]$
\EndFor
\item Return $P$
\end{algorithmic}
\end{algorithm}

\subsubsection{Fitness Evaluation} 
\begin{figure}[!ht]
\centering
\includegraphics[width=\columnwidth]{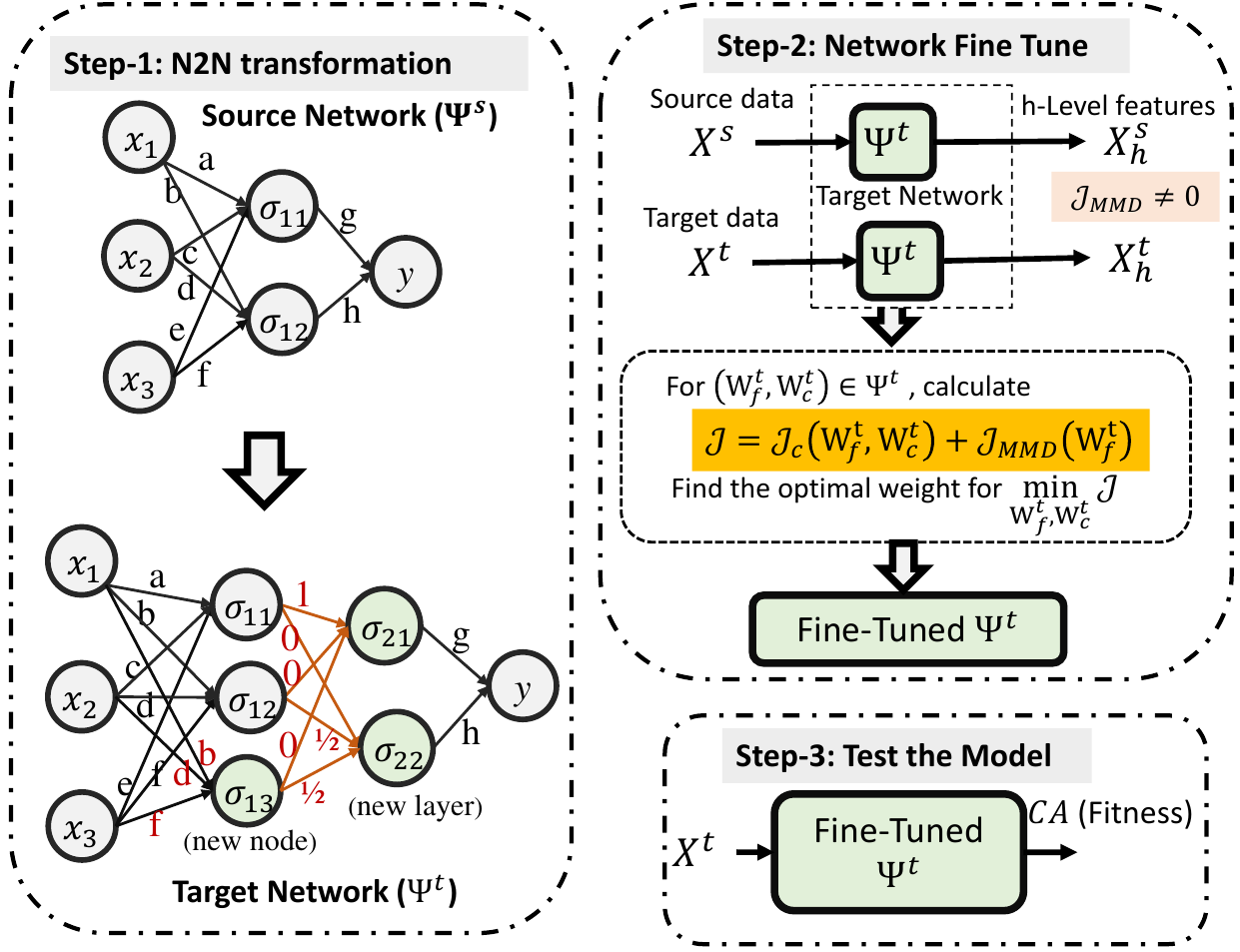}
\caption{Fitness evaluation strategy.}
\label{fig:evafit}
\end{figure}

For the quick fitness evaluation of individuals (DNN models), we adopt the strategy of quick learning mechanism proposed in \cite{aks_quick}. The first step of the fitness evaluation strategy is the network transformation, as illustrated in Fig. \ref{fig:evafit}(a). Since at the first generation, the initial (source) model trained on the source data has been transformed, the cost function for fine-tuning should include classification loss ($\mathcal{J}_c$) as well as the MMD term ($\mathcal{J}_{MMD}$) \cite{aks_quick}. For the given $C$ class problem, $\mathcal{J}_c$ and $\mathcal{J}_{MMD}$ are defined as 
\begin{equation}
\mathcal{J}_{c} = \frac{1}{N^{s}}\l[\sum_{j=1}^{N^{s}}\sum_{i=1}^{C}I[y_j = C]\log{\frac{e^{(w_i^Tf(x_j^{s})+b_i)}}{\sum_{i=1}^C e^{(w_i^Tf(x_j^{s})+b_i)}}}\r]
\end{equation}
\begin{equation}
{{\mathcal J}_{MMD}} = \sum_{i=1}^{C}{\left\Vert {\frac{1}{N_i^{s}}\sum \limits_{p = 1}^{N_i^{s}} {f (x_{i,p}^{s})} - \frac{1}{N_i^{t}}\sum \limits_{q = 1}^{N_i^{t}} {f (x_{i,q}^{t})}} \right\Vert _{\mathcal H}^2}
\end{equation}
where, $f(.)=$ h-level features output of DNN, $N_i^{s}$ and $N_i^{t}$ are the number of samples in the $i^{th}$ class of ${\rm X^{s}}$ and ${\rm X^{t}}$ respectively. $y_j$ be the source label and $[w_i,\, b_i]$ be the weight and bias connecting $i^{th}$ node in the output layer (softmax). 

Let $W_f^t\in\Psi^t\; \&\; W_c^t\in\Psi^t$ be the weight matrices of the DNN feature extractor and the softmax classifier, respectively, then the cost function minimization objective can be expressed as in Eq.(\ref{eq:cost}). 
\begin{align}
    \min_{W^t_f, W^t_c}\mathcal{J} = \min_{W^t_f, W^t_c}\l[\mathcal{J}_{c}(W^t_f, W^t_c)  + \gamma\mathcal{J}_{MMD}{(W^t_f)}\r]
    \label{eq:cost}
\end{align}
where $\gamma$ is a positive fractional value and represents the trade-off between the regularization term and the MMD term. After training, classification accuracy ($CA$) for all the models in the population ($P$) is evaluated and stored as fitness vector $P_{fit}$.
\begin{algorithm}[!ht]
\caption{Fitness Evaluation for EvoN2N}\label{algo:evafit}
\begin{algorithmic}[1]
\item $Inputs \longleftarrow (P,\; \Psi^s,\; \mathcal{D}^s,\; \mathcal{D}^t)$.
\item $N_p\longleftarrow$ population size \hspace{0.5cm} //number of individuals in $P$.
\For{$p \,= \,1 \,: \,N_p$}
\item $\Psi^t\longleftarrow\textrm{Net2Net}(\Psi^s)$ //Transform the source network to target network ($p^{th}$ model in $P$) as depicted in Fig. \ref{fig:evafit}.
\item Fine-tune the target network ($\Psi^t$) to solve Eq. (\ref{eq:cost}).
\item $P_{fit}(p) \longleftarrow \textrm{classification accuracy }(CA)$ of the final network on the validation data.
\EndFor
\item $\Psi^t_{best}\longleftarrow \textrm{Best model}$ \hspace{0.1cm} // Find the model with the maximum $CA$ and minimum number of model parameters.
\item Return $P_{fit},\; \Psi^t_{best}$
\end{algorithmic}
\end{algorithm}

\subsubsection{Parent Selection using Rank and Crowded Distance}\label{sec:p_selection} Selection of parents is required to create the population for the next generation. Algorithm \ref{algo:tour} presents the parent selection procedure assuming that each individual in the combined population $R = (P\cup Q)$ is assigned with (i) non-dominating rank ($r_p\in R_{rank}$) and (ii) crowding distance ($d_p\in R_{crowd}$).
\begin{algorithm}[!ht]
\caption{Parent Selection using $R_{rank}$ and $R_{crowd}$} \label{algo:tour}
\begin{algorithmic}[1]
\item $Inputs\longleftarrow (R, R_{crowd}, R_{rank})$
\item $N_{pf}\longleftarrow \textrm{length}(\textrm{unique}(R_{rank}))$ \hspace{0.1cm} //Maximum number of possible Pareto-front.
\item $pf\longleftarrow 1$ //set Pareto-front at 1
\item $p\longleftarrow 0$ \hspace{1cm} //solution counter $p$ at zero.
\While{$pf\leq N_{pf}$}
    \If{$p+\sum(R_{rank} == pf) \leq N_p$}{
    
       {$n = \sum(R_{rank} == pf)\}$;
       
        $P\{p+1:p+n = R\{R_{rank} == pf\}$;
        
        $p = p+\sum(R_{rank} == pf)$};
    \Else
    
        {$q = N_p - p$; //number of rest of the members in $P$.
        
        $qf= R\{R_{rank}==pf\}$; //The rest members in $P$
        
        $d_{qf} = R_{crowd}(R_{rank}==pf)$; // $d$ = distance
        
        $index = sort(d_{qf}, ``descend")$;
        
        $qf = qf(index)$;
        
        $P\{p+1:p+q\}=qf(1:p)$;
        
        $p=p+q$};}
    \EndIf\\
    $pf = pf+1$
\EndWhile
\item Return $P$
\end{algorithmic}
\end{algorithm}

\subsubsection{Crossover and Mutation}\label{sec:cross_mut} The processes of crossover and mutation are required for local search and global search, respectively, for optimal search. Due to the variable length of chromosomes, crossover is one of the major challenges for the evolution of DNN architecture. The proposed crossover method has two steps: (i) single-point depth crossover for depth variation and (ii) common depth simulated binary crossover for gene variation. The crossover process is illustrated in Fig. \ref{fig:cross}. The offspring generation by crossover and mutation is summarized in Algorithm \ref{algo:cross}:
\begin{figure*}[h!]
\centering
\includegraphics[width=\textwidth]{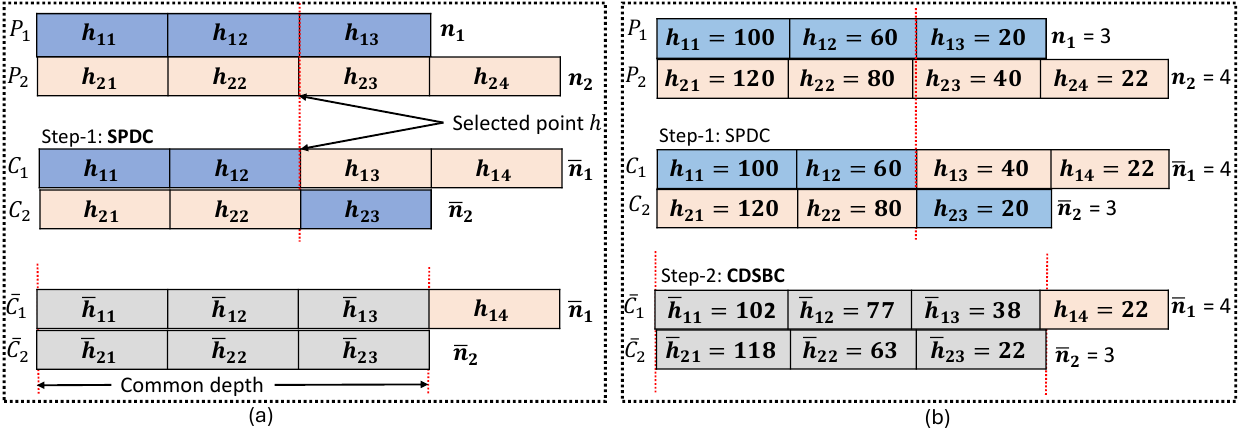}
\caption{Crossover of two different length chromosomes: (a) $n_i$ = number of hidden layers in $i^{th}$ chromosome (model), $h_{ij}$ = number of hidden nodes in $j^{th}$ hidden layer of $i^{th}$ model, $\bar{n}_i$ = number of hidden layers in $i^{th}$ offspring model, and $\bar{h}_{ij}$ = number of hidden nodes in $j^{th}$ hidden layer of $i^{th}$ offspring model; (b) Illustration of crossover with numerical values (values are taken for illustration only).}
\label{fig:cross}
\end{figure*}
\begin{algorithm}[!ht]
\caption{Offspring Generation: Crossover and Mutation}\label{algo:cross}
\begin{algorithmic}[1]
\item $Inputs:$ $P$= Parent Population, $p_c$ = Crossover Probability, $p_m$ = Mutation Probability, $N_p$ = Population Size. 
\item $I_{c}=$ indices of random $p_c*100$\% members from $P$.
\While{$I_{c}$ in not empty}
\item Select $P_1 = P\{i_1\} \;\& \;P_2 = P\{i_2\}$, where, $(i_1, i_2)=$ two random indices from $I_{c}$.
\item $(\bar{C}_1, \bar{C}_2) =$ apply crossover operator on $(P_1, P_2)$. //as illustrated in Fig. \ref{fig:cross}.
\item Replace $(P\{i_1\}, P\{i_2\})$ by $(\bar{C}_1, \bar{C}_2)$.
\item Remove $i_1, i_2$ from $I_{c}$.
\EndWhile
\item $P_{m}=$ generate $N_p$ new populations using steps in 3-8.
\item $I_{m}=$ indices of random $p_m*100$\% members from $P_m$.
\item $P\{I_m\} = P_m\{I_m\}$
\item Return $P$
\end{algorithmic}
\end{algorithm}

\section{Experimental Results and Discussion}
\label{experi}
This section demonstrates the effectiveness of the proposed frameworks on bearing fault datasets and gearbox fault datasets under variable working/load conditions. The bearing fault datasets are taken from (i)  CWRU fault diagnosis bearing data \cite{cwru} and (ii) Paderborn University (PBU) dataset \cite{paderborn}. The gearbox fault detection (GFD) dataset \cite{gfd}.

\subsection{Setup Description} 
\subsubsection{CWRU Bearing Data \cite{cwru}} \label{sub:cwru_data} The CWRU dataset provided by Case Western Reserve University (CWRU) Bearing Data Center contains three types of faults (inner raceway: \textbf{IR}, rolling element (i.e. ball): \textbf{B}, and outer raceway: \textbf{OR}) on the drive-end (DE) and fan-end (FE) bearings. The details of the experimental setup of the bearing test rig can be found in \cite{cwru}. Faults were artificially seeded on the bearings with fault diameters ranging from 0.007 to 0.028 inches (7 to 28 mil) using the electro-discharge machining process. The vibration signals were recorded under various operating conditions (motor loads 0, 1, 2, \& 3 hp and motor speed varying from 1730 to 1797 RPM).

\subsubsection{PBU Dataset \cite{paderborn}} \label{sub:paderborn_data}
C. Lessmeier \textit{et. al.} \cite{paderborn} provided a benchmark dataset for condition-based monitoring of electrical rotating machines running under a wide variety of motor load and rotational speed conditions. 
The machine is operated in four settings of load and rotational speed (abbreviated as L.S.). These are  (i) \textbf{LS1: N09\_M07\_F10} (speed = 900 rpm, torque = 0.7 Nm \& radial force = 1000 N) (ii) \textbf{LS2:  N15\_M01\_F10} (speed = 1500 rpm, torque = 0.1 Nm \& radial force = 1000 N) (iii) \textbf{LS3: N15\_M07\_F04} (speed = 1500 rpm, torque = 0.7 Nm \& radial force = 400 N) and (iv) \textbf{LS4: N15\_M07\_F10} (speed = 1500 rpm, torque = 0.7 Nm \& radial force = 1000 N). Experiments were conducted on 32 different bearings: 6 healthy bearings, 12 artificially damaged bearings, and 14 bearings damaged by accelerated lifetime tests. The dataset from each experiment contains phase current, vibration signal, radial forces,  torque, and bearing temperature. From each experiment, 20 measurements, each of 4 seconds, were taken for each of the load settings LS1, LS2, LS3, and LS4. The measurement data represents two faulty states of the machine: (i) inner race \textbf{(IR)} fault and (ii) outer race (\textbf{OR}) fault. Experiments were conducted with different levels of damage called the extent of damage.

\subsubsection{Gearbox Fault Detection (GFD) Dataset \cite{gfd}} \label{sub:GFD} A SpectraQuest’s Gearbox Fault Diagnostics Simulator has been used to record the gear vibration signals. The dataset contains four vibration signals recorded using four sensors installed in four different directions. The recorded signals represent two different states of gear conditions, namely healthy/normal (\textbf{N}) and broken tooth (\textbf{BT}). The dataset for each state of gear fault contains 10 files of vibration signals recorded under various load conditions from 0 to 90\%.

\subsection{Data Processing and Evaluation Scheme}\label{e.s.} First, the recorded time-series data is segmented and scaled down to [0, 1] scale by using the min-max normalization technique: $x_n = (x-x_{min})/(x_{max}-x_{min})$, where $x_{min}$ and $x_{max}$ are the minimum and maximum values of the dataset $x$. A large number of data points from the time series signal are converted to samples of reduced dimensionality by the segmentation process. For CWRU, the time series signal is segmented with a segment length of 100 data points. For example, a signal with a length of 121200 points is converted into $1212\times100$ samples. Similarly, the GFD and PBU data are segmented using a segment length of 400 data points.

Here, source and target datasets are formed as discussed in the following cases.\\
\textit{\textbf{Case-1:}} \textbf{CWRU dataset:}
\begin{enumerate}[i)]
    \item \textbf{Source Data:} The time-series signals recorded at \textit{12 Hz DE} with a motor load of $0$ hp and speed of $1796$ RPM are segmented to create $1200\times100$ samples per class. After merging the segmented dataset from each class (\textbf{N}, \textbf{IR}, \textbf{OR}, and \textbf{B}), it contains a total of $4800\times100$ samples.
    \item \textbf{Target-1:} The time-series signals recorded at \textit{12k Hz FE} with motor loads of $1$, $2$, \& $3$ hp, fault diameter (F.D.) varying from $7$ to $21$ mil, and speed of $1772$ RPM are used to prepare the dataset for target-1 (\textbf{T1}) under nine different cases with F.D.'s and motor loads as shown in table \ref{Table: CWRU}. 
\end{enumerate}
\textit{\textbf{Case-2:}} \textbf{PBU dataset:}
\begin{enumerate}[i)]
    \item \textbf{Source Data:} The time-series signal recorded for an artificially damaged bearing fault with the motor running under the LS of L1 is used to prepare the segmented samples of size $5000\times400$ from 10 files per class. 
    \item \textbf{Target-2 \& 3:}  Two different sets of target datasets  (target-2 (\textbf{T2}) \& target-3 (\textbf{T3})) are created by considering different levels of damage (extent of damage) under four L.S.'s of LS1, LS2, LS3, \& LS4  as given in table \ref{Table: paderborn}.
\end{enumerate}
\textit{\textbf{Case-3:}} \textbf{Gearbox fault dataset:} 
\begin{enumerate}[i)]
    \item \textbf{Source Data:} $80000$ sample points from the file corresponding to zero load and the sensor '1' is segmented to convert into $800\times100$ samples per class.
    \item \textbf{Target-4:} Target-4 (T4) is created under three different load conditions using the signals from the files corresponding to 30, 60, \&90 \% of load as summarised in Table \ref{Table: GFD_results}. For each load, a small number of sample points $30000$ are used to create a segmented data set of $300\times 100$ samples per class.
\end{enumerate}

Each of the target datasets is split into train, test, and validate datasets using the random sampling method. $20\%$ of the available target samples are kept for testing, $16\%$ for validation, and $64\%$ for training.

\begin{table*}[ht]
\centering %
\caption{\small  CA for CWRU bearing fault data (T1) recorded at Fan End (FE) with three fault diameters (F.D.s)} %
\resizebox{\textwidth}{!}{%
\begin{tabular}{ |c|c|c|c|c|c|c|c|c|c|c|c|}
\hline
    \multirow{2}{*}{\makecell{Target}}  &
    \multirow{2}{*}{\makecell{F.D.}}  &  \multirow{2}{*} {\makecell{Load}}  & \multirow{2}{*}{\makecell{SVM\\ \cite{svm}}} &  \multirow{2}{*}{\makecell{DNN\\ \cite{sparseAE}}} &  \multirow{2}{*}{\makecell{DAGCN\\ \cite{dagcn}}} &  \multicolumn{2}{c|} {N2N \cite{aks_quick}} &  \multirow{2}{*} {\makecell{EvoDCNN\\ \cite{EvoDCNN}}} & \multirow{2}{*} {\makecell{DenseNAS\\ \cite{c1rv1}}} &  \multirow{2}{*} {\makecell{{WGANML}\\ \cite{newfd1}}} & \multirow{2}{*} {\textbf{EvoN2N}}  \\
\cline{7-8}
    & & & & & & \makecell{W. D. A.} & \makecell{D. A.} &  &  &  & \\
\hline
\multirow{9}{*}{T1} & \multirow{3}{*}{\makecell{FE\\7 mil}} & 1hp & 87.5  & 96.7  & 98.9  & 98.9  & 98.9  & 99.6  & 99.6  & 99.6  & \textbf{100.0} \\
\cline{3-12} && 2hp &  98.1  & 95.9  & 90.3  & 97.1  & 98.1  & 99.6  & 99.6  & 99.6  & \textbf{100.0} \\
\cline{3-12} &&3hp  &  93.8  & 98.8  & 98.4  & 99.4  & 99.4  & 99.6  & 99.4  & 99.6  & \textbf{100.0} \\
\cline{2-12}
& \multirow{3}{*}{\makecell{FE\\14 mil}}&  1hp  & 90.3  & 94.8  & 95.6  & 99.2  & 99.2  & 99.6  & 99.2  & 99.6  & \textbf{100.0} \\
\cline{3-12} && 2hp & 98.4  & 97.7  & 98.7  & 97.7  & 98.7  & 98.1  & 98.7  & 98.4  & \textbf{99.2} \\
\cline{3-12} && 3hp & 98.4  & 96.9  & 97.7  & 99.2  & 98.6  & 98.4  & 99.2  & 98.6  & \textbf{99.6} \\
\cline{2-12}
& \multirow{3}{*}{\makecell{FE\\21 mil}} &  1hp & 96.3  & 95.6  & 96.9  & 95.6  & 96.6  & 93.8  & 96.6  & 96.6  & \textbf{98.8} \\
\cline{3-12} && 2hp & 88.8  & 90.7  & 90.7  & 90.7  & 90.7  & 90.1  & 93.8  & 90.7  & \textbf{95.4} \\
\cline{3-12} && 3hp & 91.1  & 93.1  & 93.8  & 91.1  & 93.8  & 92.8  & 92.8  & 92.8  & \textbf{95.8} \\
\hline
\rowcolor{Gray}\multicolumn{3}{|c|} {Standard Deviation} & 4.4   & 2.5   & 3.4   & 3.4   & 3.0   & 3.6   & 2.6   & 3.3   & \textbf{1.8 }\\
\hline 
\end{tabular}}
\label{Table: CWRU}
\end{table*}

\begin{table*}[ht]
\centering %
\caption{\textsc{\small  CA for PBU Dataset (T2 \& T3) with different levels of fault and different load settings (L.S.)}} %
\resizebox{\textwidth}{!}{%
\begin{tabular}{ |c|c|c|c|c|c|c|c|c|c|c|}
\hline
    \multirow{2}{*}{\makecell{Target}}  &  \multirow{2}{*} {\makecell{L.S.}}  & \multirow{2}{*}{\makecell{SVM\\ \cite{svm}}} &  \multirow{2}{*}{\makecell{DNN\\ \cite{sparseAE}}} &  \multirow{2}{*}{\makecell{DAGCN\\ \cite{dagcn}}} &  \multicolumn{2}{c|} {N2N \cite{aks_quick}} &  \multirow{2}{*} {\makecell{EvoDCNN\\ \cite{EvoDCNN}}} & \multirow{2}{*} {\makecell{DenseNAS\\ \cite{c1rv1}}} &  \multirow{2}{*} {\makecell{{WGANML}\\ \cite{newfd1}}} & \multirow{2}{*} {\textbf{EvoN2N}}  \\
\cline{6-7}
    & & & & & \makecell{W. D. A.} & \makecell{D. A.} &  &  &   &  \\
\hline
 \multirow{4}{*}{T2} & L1  &  96.3  & 97.9  & 97.2  & 99.6  & 99.6  & 100.0 & 99.9  & 99.2  & \textbf{100.0} \\
\cline{2-11} & L2 &  90.0  & 94.6  & 98.6  & 95.1  & 96.1  & 99.6  & 99.6  & 97.5  & \textbf{99.8} \\
\cline{2-11} & L3  &  89.2  & 92.9  & 94.4  & 94.4  & 94.4  & 97.5  & 94.4  & 97.5  & \textbf{97.7} \\
\cline{2-11} & L4  &  89.2  & 93.8  & 95.3  & 97.2  & 95.3  & 99.8  & 99.6  & 97.2  & \textbf{100.0} \\
\hline
\multirow{3}{*}{T3}&  L1  &  97.9  & 98.6  & 93.8  & 98.6  & 99.2  & 99.6  & 99.2  & 99.2  & \textbf{99.8} \\
\cline{2-11} & L2  &  95.8  & 97.9  & 94.2  & 96.3  & 96.6  & 98.6  & 98.6  & 99.2  & \textbf{100.0} \\
\cline{2-11} &L3  &  95.8  & 94.2  & 97.9  & 95.7  & 96.8  & 97.2  & 96.8  & 99.6  & \textbf{99.8} \\
\cline{2-11} & L4  &  93.3  & 93.3  & 95.7  & 95.7  & 95.7  & 93.3  & 96.8  & 98.6  & \textbf{99.2} \\
\hline
\rowcolor{Gray}\multicolumn{2}{|c|} {Standard Deviation} &  3.5   & 2.3   & 1.8   & 1.8   & 1.8   & 2.2   & 1.9   & 1.0   & \textbf{0.8} \\
 \hline
\end{tabular}}
\label{Table: paderborn}
\end{table*}

\begin{table*}[ht]
\centering %
\caption{\textsc{\small  CA for Gearbox Fault Detection Dataset (T4) with 30, 50, 70, 90\% of load}} %
\resizebox{\textwidth}{!}{%
\begin{tabular}{ |c|c|c|c|c|c|c|c|c|c|c|}
\hline
    \multirow{2}{*}{\makecell{Target}}  &  \multirow{2}{*} {\makecell{Load\\ (\%)}}  & \multirow{2}{*}{\makecell{SVM\\ \cite{svm}}} &  \multirow{2}{*}{\makecell{DNN\\ \cite{sparseAE}}} &  \multirow{2}{*}{\makecell{DAGCN\\ \cite{dagcn}}} &  \multicolumn{2}{c|} {N2N \cite{aks_quick}} &  \multirow{2}{*} {\makecell{EvoDCNN\\ \cite{EvoDCNN}}} & \multirow{2}{*} {\makecell{DenseNAS\\ \cite{c1rv1}}} &  \multirow{2}{*} {\makecell{{WGANML}\\ \cite{newfd1}}} & \multirow{2}{*} {\textbf{EvoN2N}}  \\
\cline{6-7}
    & & & & & \makecell{W. D. A.} & \makecell{D. A.} &  &  &   &  \\
\hline
 \multirow{4}{*}{T4} & 30.0 & 83.3  & 90    & 92.5  & 95.0  & 95.0  & 98.3  & 95.8  & 99.2  & \textbf{100.0} \\
\cline{2-11} & 50.0 & 87.5  & 90.8  & 93.3  & 93.3  & 95.8  & 95.8  & 98.3  & 98.3  & \textbf{99.2} \\  
\cline{2-11} & 70.0 & 87.5  & 95    & 93.3  & 92.5  & 93.3  & 95.0  & 95.8  & 99.2  & \textbf{99.8} \\
\cline{2-11} & 90.0 & 89.2  & 93.3  & 95    & 93.3  & 95.0  & 98.3  & 99.2  & 98.3  & \textbf{99.6} \\
\hline
\rowcolor{Gray}\multicolumn{2}{|c|} {Standard Deviation} & 2.5   & 2.3   & 1.1   & 1.1   & 1.1   & 1.7   & 1.7   & 0.5   & \textbf{0.3} \\ 
 \hline
\end{tabular}}
\label{Table: GFD_results}
\end{table*}

\subsection{Evaluation metrics} Classification performance of a diagnostic model is measured in terms of \textbf{ classification precision} ($CA$) as widely accepted in the literature \cite{longRTRL, lwen}.
\begin{align}
    CA = \frac{\left\vert{{\bf{x}}:{\bf{x}}\in{{ X}^{te}} \,\wedge y={\cal F}\left({\bf{x}}\right)}\right\vert}{\left\vert{{\bf{x}}:{\bf{x}}\in{{X}_{te}}}\right\vert}\times 100\%
\end{align}
where, ${X}_{te}$ is the test data, ${\cal F}\left({\bf{x}}\right)$ and $y$ are the predicted and the true labels. Additionally, the improvement analysis of the proposed method with respect to a baseline method is described in terms of transfer improvement ($TI$). $TI$ is calculated as $TI = \overline{CA} - \overline{CA}_{baseline}$, where, $\overline{CA}$ is the average $CA$ for the dataset under various operating conditions.

\begin{figure}[!ht]
	\centering
	\includegraphics[width=\columnwidth]{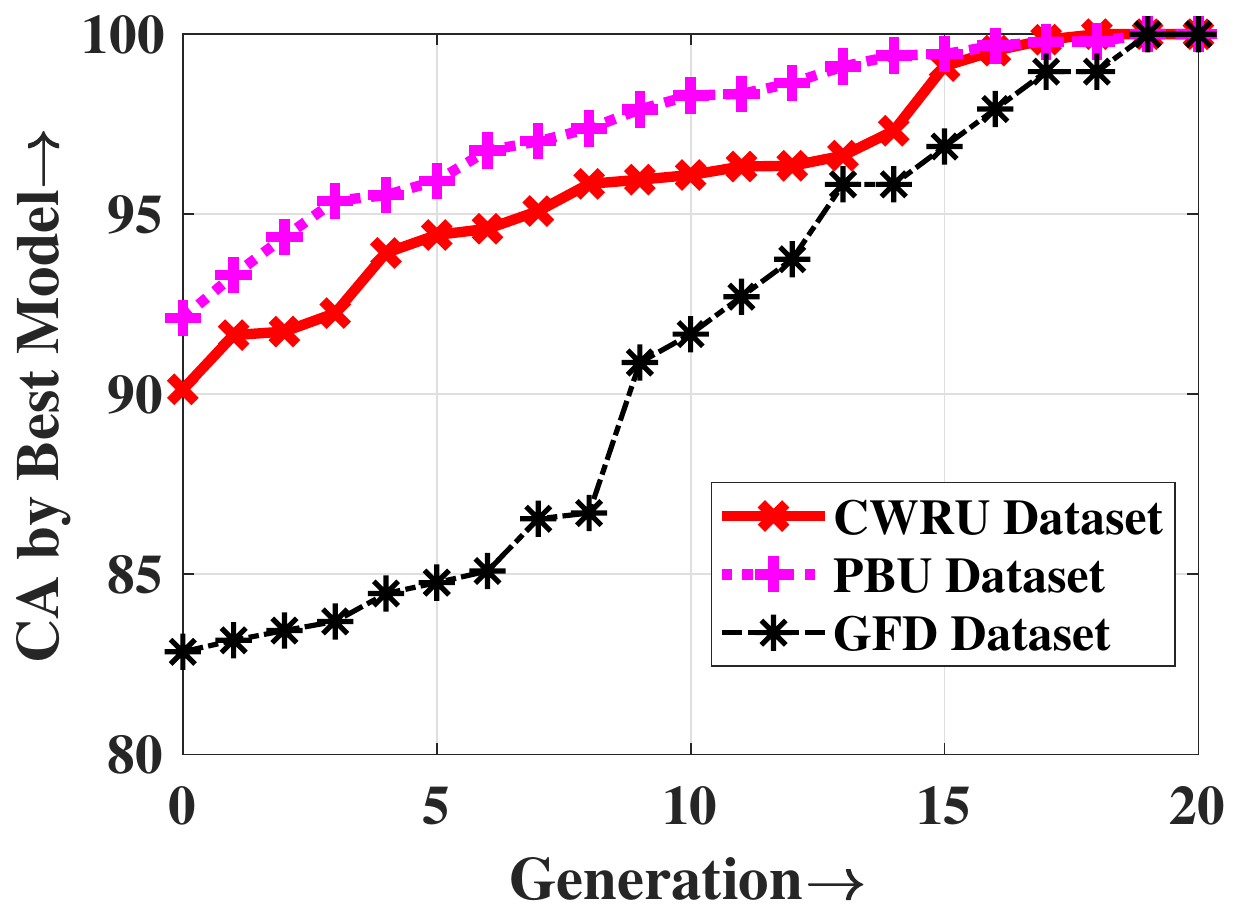}
	\caption{Growth of CA with generation for the first load condition from each case.}
    \label{fig:acc_curve}
\end{figure}

\subsection{Implementation and Training Parameters}  The proposed EvoN2N as described in Algorithm \ref{algo:proposed} requires an initial DNN model ($\Psi^s$) termed as the teacher model or the source model. Therefore, first, a source model with architecture (we choose a DNN model with $3$ hidden layers $80-40-20$) is trained on the source dataset (as mentioned in Section \ref{e.s.} for three cases) using back-propagation gradient descent. 

Now, with this source network, the proposed framework \ref{algo:proposed} is applied to the target datasets T1 to T4. Initial parameters for the evolutionary algorithm are chosen: (i) population size ($N_p$) = 100, (ii) probability of crossover ($P_c$) = 0.8, (iii) probability of mutation ($P_m$) = 0.2, and (iv) the number of maximum generation = 20. The maximum ranges for the number of hidden layers and nodes are selected as $n_h\in[1, \;8]$ and $h\in[4, \;400]$ respectively. The best models obtained for each of the cases of the target dataset (T1, T2, T3 \& T4) are tested on the corresponding test data. Performances in terms of $CA$ are tabulated in tables \ref{Table: CWRU}, \ref{Table: paderborn} \& \ref{Table: GFD_results}.

\begin{figure}[!ht]
	\centering
	\includegraphics[width=\columnwidth]{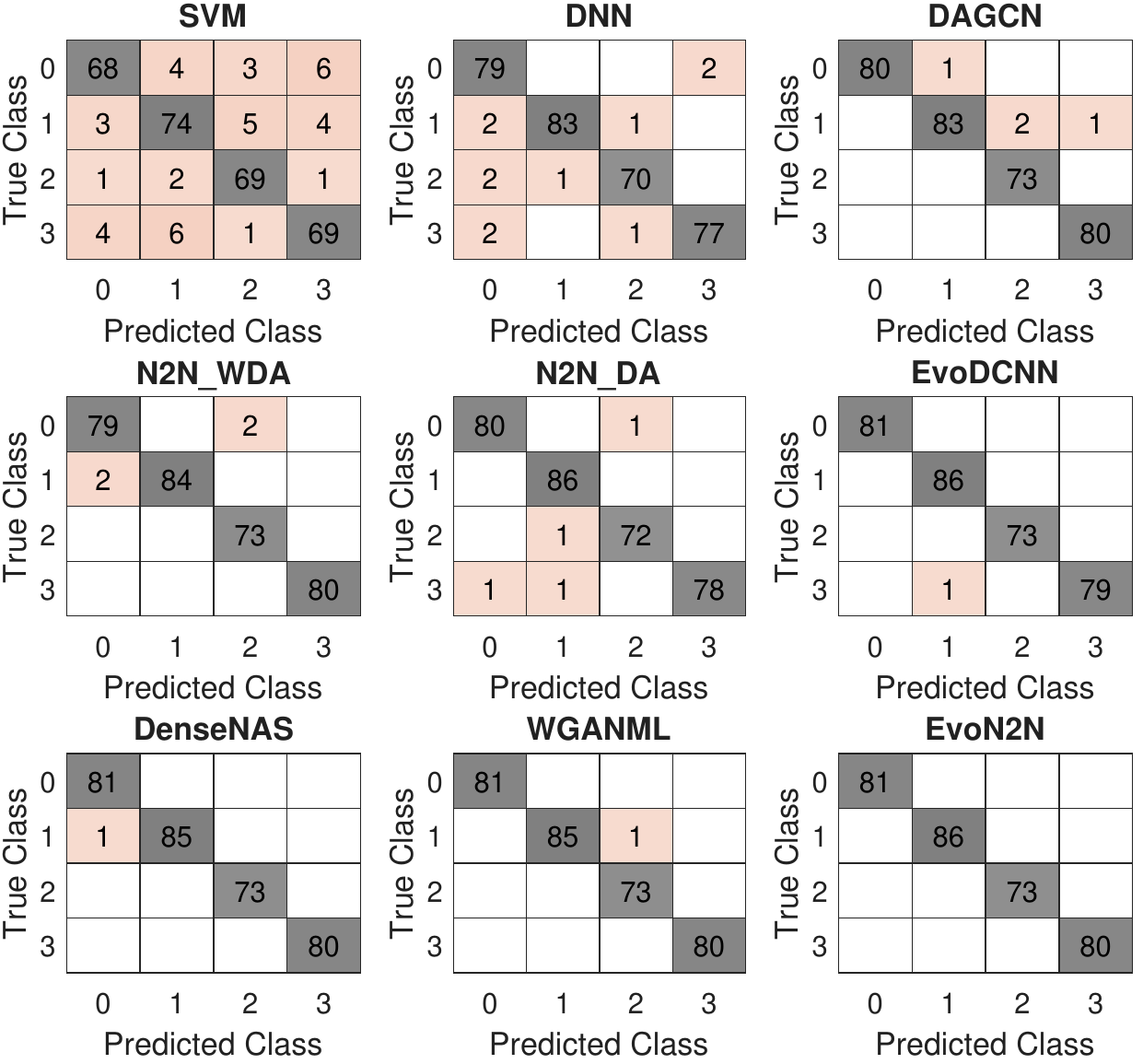}
	\caption{Confusion matrix for T1 (CWRU, 7mil, 1hp): class labels \{`0', `1', `2', `3'\} represents the class names \{`H', `IR', `B', `OR'\}.}
    \label{fig:cm}
\end{figure}
\begin{figure*}[!ht]
	\centering
	\includegraphics[width=\textwidth]{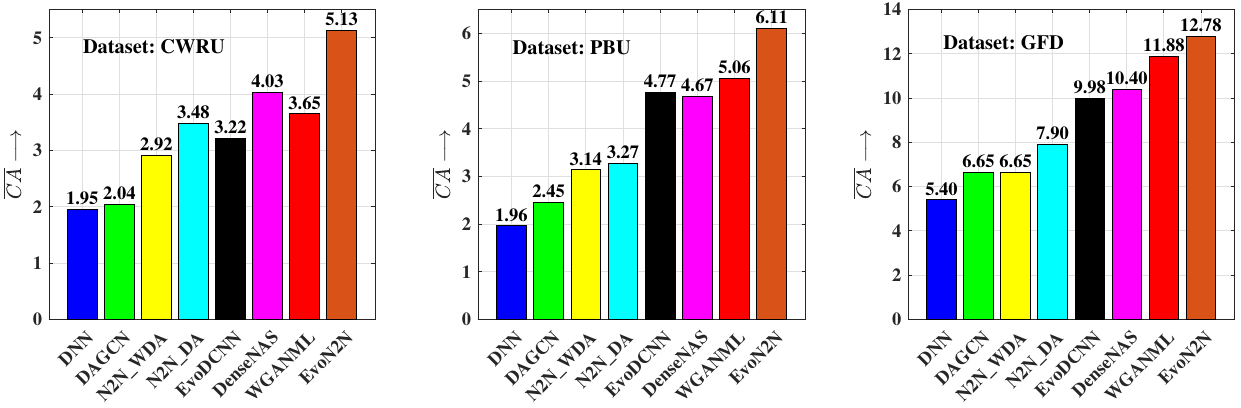}
	\caption{Transfer Improvement (TI) of average $CA$ ($\overline{CA}$) on (i) CWRU dataset, (ii) PBU dataset, and (iii) GFD dataset, assuming SVM be the baseline method.}
    \label{fig:TI}
\end{figure*}

The performance of the proposed framework is compared with the state-of-the-art method most popularly used for intelligent fault diagnosis. The selected algorithms are support vector machines (SVM): the baseline method \cite{svm}, deep neural network (DNN) \cite{sparseAE}, Domain Adversarial Graph Convolutional Network (DAGCN) \cite{dagcn}, Net2Net without domain adaptation (N2N\_WDA) \cite{aks_quick}, Net2Net with domain adaptation (N2N\_DA) \cite{aks_quick}, evolutionary deep CNN (EvoDCNN) \cite{EvoDCNN}, DenseNAS \cite{c1rv1}, and Wasserstein GAN meta-learning (WGANML) \cite{newfd1}. The architecture for DNN is kept the same as the source network. The model architectures and hyperparameters for other models are chosen based on standard values provided in their references. The population size and the maximum number of generations for EvoDCNN are kept the same as the proposed method. All these models are trained on the same data as described in section \ref{e.s.} using the methods suggested in the cited references. The classification accuracies are tabulated in tables \ref{Table: CWRU} and \ref{Table: paderborn}. For better visualization of the classification performance, the confusion matrix for one of the cases taken from T1 (7mil, 1hp) has been shown in Fig. \ref{fig:cm}.  For comparative analysis, Fig. \ref{fig:TI} shows the transfer improvement ($TIs$) in terms of average $CA$ ($\overline{CA}$) for the three cases: (i) CWRU dataset, (ii) PBU dataset, and (iii) GFD dataset.

\subsection{Ablation Study:} The main contribution of this research work is the evolutionary network-to-network search with the NSGA-II framework. The model architecture is encoded using variable-length chromosomes to exploit the search space of the number of layers and nodes. The training process for the fitness evaluation involves model initialization using N2N transformation \cite{aks_quick}, followed by fine-tuning with a small number of target samples.  The ablation study to show the impact of the key parameters is presented as follows. 
\subsubsection{\textbf{Effect of N2N-based quick learning}} 
If the individuals in the populations are trained and evaluated from scratch at each generation, the DNN model parameters for each chromosome must be initialized with random initial values. The average time taken for one generation with a population size of 100 is recorded to be 20584 seconds using random initialization of model parameters, while the average time for one generation with the same population size and the computational machine was recorded to be 9500 seconds using the N2N-based quick learning mechanism.

\subsubsection{\textbf{Effect of population size}}
Table \ref{tab:evoArch}. shows the optimal architecture obtained at various generations for the population size of 50 and 100. It can be seen that the proposed framework does not converse fully in 20 generations with a population size of 50, however, it converses faster with the larger population size.
\begin{table}[!ht]
  \centering
  \caption{Evolution of best architecture at every $5^{th}$ generation (\textbf{CWRU FE, 7 mil FD, 1 hp load})}
  \resizebox{\columnwidth}{!}{%
    \begin{tabular}{|c|l|c|l|c|}
    \hline
    \multirow{2}[4]{*}{\textbf{\#Gen}}   & \multicolumn{2}{c|}{\textbf{Pop. Size = 50}} & \multicolumn{2}{c|}{\textbf{Pop. Size = 100}} \\
\cline{2-5}          
 & \textbf{Best Arch.} & \textbf{\% CA} & \textbf{Best Arch.} & \textbf{\% CA} \\
    \hline
    0     & 100-20 & 89.8   & 90-88-73 & 90 \\
    \hline
    5     & 98-95-61 & 94.3  & 100-95-75-73-64-35 & 94.4 \\
    \hline
    10    & 100-76-66 & 95.9  & 100-45 & 96.1  \\
    \hline
    15    & 99-86-81-74 & 98.1  & 97-75-75-60 & 99.1 \\
    \hline
    20    & {100-96-84-63} & {99.7}   & \textbf{100-100-80-73} & \textbf{100} \\
    \hline
    \end{tabular}}
  \label{tab:evoArch}%
\end{table}%

\subsubsection{\textbf{Effect of number of generation}} From Table \ref{tab:evoArch} and the accuracy curve shown in Fig. \ref{fig:acc_curve}, it can be observed that the algorithm takes about 20 generations to reach to its optimal for each of the dataset cases.

\subsection{Hypothesis Testing} \label{sec:hypotesting} To show the statistical significance of the proposed EvoN2N framework, hypothesis testing is performed using the Friedman test and the Bonferroni-Dunn (BD) test on the results shown for target datasets T1 to T4. The Friedman test is performed to check the null hypothesis that all classifiers (methods) are the same against the alternative hypothesis of all the classifiers being significantly different. If the null hypothesis is rejected, the BD test is performed as the post-hoc test to obtain an alternative hypothesis. The Friedman test and the BD test are described as follows:

\begin{table*}[!ht]
  \centering
  \caption{Rank table for results shown in Tables \ref{Table: CWRU}, \ref{Table: paderborn}, and \ref{Table: GFD_results} for all datasets T1-T4.}
  \resizebox{\textwidth}{!}{%
    \begin{tabular}{|c|c|c|c|c|c|c|c|c|c|c|}
    \hline
    {\multirow{2}{*}{{Target-load}}} & \multirow{2}{*}{\makecell{SVM\\ \cite{svm}}} &  \multirow{2}{*}{\makecell{DNN\\ \cite{sparseAE}}} &  \multirow{2}{*}{\makecell{DAGCN\\ \cite{dagcn}}} &  \multicolumn{2}{c|} {N2N \cite{aks_quick}} &  \multirow{2}{*} {\makecell{EvoDCNN\\ \cite{EvoDCNN}}} & \multirow{2}{*} {\makecell{DenseNAS\\ \cite{c1rv1}}} &  \multirow{2}{*} {\makecell{{WGANML}\\ \cite{newfd1}}} & \multirow{2}{*} {EvoN2N}  \\
\cline{5-6}
    & & & & \makecell{W. D. A.} & \makecell{D. A.} &  &  &  & \\
    \hline
    Average ranks & 8.1   & 7.3   &   6.0    & 6.2   & 5.1   & 4.2   & 3.5   & 3.6   & 1.0 \\
    \hline
    \end{tabular}}%
  \label{ch4:rank1}%
\end{table*}

\textbf{Friedman test:} The Friedman test \cite{friedmann1940} is a non-parametric extension of two-way analysis of variance (ANOVA) to test the column effect. It compares the average rank of the classifiers among all the datasets in terms of $\chi^2$ and Friedman statistics $S_F$. $\chi^2$ and $S_F$ are calculated using the following equations: 
\begin{equation}\label{eq:fs1}
\chi^2 = \frac{12N_t}{N_m(N_m+1)}\l[\sum_{i}^{N_m}\mathcal{R}_i^2\; - \; \frac{N_m(N_m+1)^2}{4}\r]
\end{equation}
\begin{equation}\label{eq:fs2}
    S_F = \frac{(N_t-1)\chi^2}{N_t(N_m-1)-\chi^2}
\end{equation}
where $N_t$ is the number of sub-cases (datasets) in the target data domain, $N_m$ is the number of methods, and  $\mathcal{R}_i$ represents the average rank of the $i^{th}$ method on the sub-cases. $S_F$ is distributed according to $(N_m-1)$ and $(N_m-1)(N_t-1)$ degree of freedom ($d_F$). The critical values for the Friedman statistics distributions can be found in Tables A4 and A10, provided as the appendix in the book \cite{sheskin2003handbook}. If the Friedman statistical parameter $S_F$ is below the critical value, the null hypothesis is accepted, stating that all the algorithms are similar.

\textbf{Bonferroni-Dunn (BD) test:} If the null hypothesis based on the Friedman test is rejected, the BD test \cite{bdtest} is followed to analyze the statistical significance of the algorithms. The BD test states that if the difference in the average rank is larger than the critical difference, the algorithm is statistically superior to the other. The critical difference ($CD$) is calculated for a given significance value ($\alpha$) according to equation (\ref{eq:cd}).
\begin{equation}\label{eq:cd}
    CD = q_{\alpha}\sqrt{\frac{N_m(N_m+1)}{6N_t}}
\end{equation}
where, $q_{\alpha}$ represents the critical values and can be found in Table 5(b) of \cite{stat_cd}.

For the hypothesis test, a combined rank table for results shown in Tables \ref{Table: CWRU}, \ref{Table: paderborn}, and \ref{Table: GFD_results} is created, which constitute a total of 21 sub-cases of datasets and 9 classifiers (methods). The average ranks of all the algorithms for the 21 datasets are calculated and shown in Table \ref{ch4:rank1}.

For the rank table shown in Table \ref{ch4:rank1}, Friedman's ANOVA table is obtained using MATLAB's function \textit{`friedman'}, and the ${p}$-value is compared on the significance level of 0.05 and 0.01. Friedman's ANOVA table is shown in Table \ref{ch4:anova1}.

\begin{table}[!ht]
  \centering
  \caption{Friedman's ANOVA Table: $S_S$ = Sum of squares and $d_F$ = degree of freedom.}
  \resizebox{\columnwidth}{!}{%
    \begin{tabular}{|c|c|c|c|c|c|}
    \hline
          Source & $S_S$ & $d_F$ & $S_S/d_F$ & $\chi^2$ & ${p}$ \\
          \hline
          Columns & 793.9048 & 8  & 99.2381 & 109.9555 & $3.89e^{-20}$ \\
          \hline
          Error & 419.0952 & 160   & 2.6193   &    -   & - \\
          \hline
          Total & 1213  & 188   &   -    &    -   & - \\
          \hline
    \end{tabular}}%
  \label{ch4:anova1}%
\end{table}%

Since the obtained \textit{p-value} = $3.89e^{-20} <0.01$, the null hypothesis is rejected. Therefore, all the methods (columns) are significantly different and the alternative hypothesis test (BD test) is followed to obtain the significance of the proposed method. 

Now, for the BD test, values of $CD$ at significance values $\alpha = 0.1$ and $\alpha = 0.05$ are calculated using equation (\ref{eq:cd}). For the number of methods $N_m = 9$, the value of $q_{\alpha}$ as obtained from Table 5(b) of \cite{stat_cd} as: $q_{0.1} = 2.498$ and $q_{0.05} = 2.724$ for significance values $\alpha = 0.1$ and $\alpha = 0.05$, respectively. Therefore, $CD_{0.1}$  = $2.111195$ and $CD_{0.05}$  = $2.3022$ . The differences in the average rank for the selected state-of-the-art methods with respect to the proposed EvoN2N framework are $7.0714$,    $6.2619$,    $4.9762$,    $5.1667$,    $4.0476$,    $3.1905$,    $2.5238$, and   $2.5476$. Rank differences are larger than $CD_{0.1}$ as well as $CD_{0.05}$. Therefore, the proposed EvoN2N is statistically significant compared to the state-of-the-art methods at the significance level of 0.05.

\subsection{Discussion} Following points can be observed from the diagnostic results shown in tables \ref{Table: CWRU}, \ref{Table: paderborn}, \& \ref{Table: GFD_results} and Fig.'s \ref{fig:cm}, \ref{fig:TI}, \& \ref{fig:acc_curve}.
\begin{enumerate}
    \item From Fig. \ref{fig:acc_curve}, it can be observed that the best model at each generation improves with evolution. Since the exploitation by the crossover and the exploration by the mutation perform both the local and the global searches, the final model obtained has the best possible DNN architecture.
    \item Fig. \ref{fig:cm} provides graphical illustration of the classification performances of different state-of-the-art methods for one of the case from T1: 7mil, 1 hp. It can be observed that the confusion matrix for the proposed framework depicts all correct classifications. 
    \item TI in Fig. \ref{fig:TI} for the three types of dataset reveals that the performance of some methods may be poorer than the baseline method SVM depending upon the dataset. However, the proposed method performs better in all cases. Therefore, the proposed evoN2N framework is an advancement for building a suitable deep learning model for the real-time scenario where dataset availability is limited.
    \item Overall, the validation results on various datasets under different natures of faults and operating conditions lead to the conclusion that the proposed framework enables us to build a reliable and accurate diagnostic model with optimal network architecture using a limited number of training samples. Also, the wide variation in the target dataset proves that the proposed framework can be beneficial for building reliable and accurate diagnostic models in real-time scenarios for the classification of real faults due to accelerated lifetime runs.
\end{enumerate}
Therefore, the proposed framework is the effective solution to ensure an accurate fault diagnosis under variable operating conditions via the evolution of the best diagnostic model.

\subsection{Complexity Analysis} The complexities of the proposed algorithm \ref{algo:proposed} are defined based on the worst complexity in each epoch. One epoch of the entire algorithm has the worst complexity, contributed by fitness evaluation, non-dominated sorting, parent selection based on crowding distance and rank, and crossover and mutation processes. Out of all these steps, fitness evaluation and non-dominated sorting have the worst complexity to contribute. 

The fitness evaluation of the DNN model involves the parameter optimization by L-BFGS with the complexity of $O(N_I*n^2)$, where $n \& N_I$ is the total number of parameters and the number of iterations required to fine-tune the DNN model. The complexity of non-dominated sorting is given by $O(MN_p^2)$, where $M\, \&\, N_p^2$ represents the number of objectives and the population size respectively. Since the number of objectives ($M$) is usually much smaller compared to the population size ($N_p$), the complexity of the proposed algorithm (EvoN2N) can be approximated as $O(N_IN_pn^2)$ where, $N_p$ be the population size, $N_I$ be the number iteration for model training in the fitness evaluation step, $n$ be the number of parameters in a DNN model.

\section{Conclusions}
\label{conclusion} This article proposed the framework of evolving Net2Net transformation for DNN architecture search. The DNN architecture to be optimized is encoded as real-coded chromosomes (individuals) in the population initialization. The best model obtained in each generation is transferred to the next generation for the initialization of individuals. Transferring the best knowledge gained makes the training of the individuals faster with the evolution. Therefore, the proposed framework has proved to be a very effective solution for fault diagnosis with a high accuracy of up to almost 100\%. Validation on various target datasets under different operating conditions justifies the suitability of the proposed method for real-time industrial applications. Also, the proposed method has a big transfer improvement over the baseline method (SVM). The proposed method does not care about the number of evolutions required to achieve the optimality. Therefore, work can further be extended for faster evolution using an agent-based population initialization (meta-reinforcement) with guided population size. The limited size of the population will accelerate the evolution process.

\ifCLASSOPTIONcaptionsoff
  \newpage
\fi

\bibliographystyle{IEEEtran.bst}
\bibliography{Reference.bib}

\begin{thebibliography}{10}
\providecommand{\url}[1]{#1}
\csname url@samestyle\endcsname
\providecommand{\newblock}{\relax}
\providecommand{\bibinfo}[2]{#2}
\providecommand{\BIBentrySTDinterwordspacing}{\spaceskip=0pt\relax}
\providecommand{\BIBentryALTinterwordstretchfactor}{4}
\providecommand{\BIBentryALTinterwordspacing}{\spaceskip=\fontdimen2\font plus
\BIBentryALTinterwordstretchfactor\fontdimen3\font minus
  \fontdimen4\font\relax}
\providecommand{\BIBforeignlanguage}[2]{{%
\expandafter\ifx\csname l@#1\endcsname\relax
\typeout{** WARNING: IEEEtran.bst: No hyphenation pattern has been}%
\typeout{** loaded for the language `#1'. Using the pattern for}%
\typeout{** the default language instead.}%
\else
\language=\csname l@#1\endcsname
\fi
#2}}
\providecommand{\BIBdecl}{\relax}
\BIBdecl

\bibitem{10504763}
Y.~Wang, J.~Shen, S.~Yang, Q.~Han, C.~Zhao, P.~Zhao, and X.~Ren, ``Knowledge
  and data dual-driven fault diagnosis in industrial scenarios: A survey,''
  \emph{IEEE Internet of Things Journal}, vol.~11, no.~11, pp.
  19\,256--19\,277, 2024.

\bibitem{HU2025102963}
C.~Hu, Z.~Zhang, C.~Li, M.~Leng, Z.~Wang, X.~Wan, and C.~Chen, ``A state of the
  art in digital twin for intelligent fault diagnosis,'' \emph{Advanced
  Engineering Informatics}, vol.~63, p. 102963, 2025.

\bibitem{10143698}
W.~Li, H.~Lan, J.~Chen, K.~Feng, and R.~Huang, ``Wavcapsnet: An interpretable
  intelligent compound fault diagnosis method by backward tracking,''
  \emph{IEEE Transactions on Instrumentation and Measurement}, vol.~72, pp.
  1--11, 2023.

\bibitem{sr2}
F.~Immovilli, \textit{et al. }, ``Diagnosis of bearing faults in induction
  machines by vibration or current signals: A critical comparison,'' \emph{IEEE
  Transactions on Industry Applications}, vol.~46, no.~4, pp. 1350--1359, 2010.

\bibitem{sr11}
W.~{Fan}, Q.~{Zhou}, J.~{Li}, and Z.~{Zhu}, ``A wavelet-based statistical
  approach for monitoring and diagnosis of compound faults with application to
  rolling bearings,'' \emph{IEEE Transactions on Automation Science and
  Engineering}, vol.~15, no.~4, pp. 1563--1572, Oct 2018.

\bibitem{rn1}
S.~Yin, S.~X. Ding, X.~Xie, and H.~Luo, ``A review on basic data-driven
  approaches for industrial process monitoring,'' \emph{IEEE Transactions on
  Industrial Electronics}, vol.~61, no.~11, pp. 6418--6428, 2014.

\bibitem{svm}
A.~Widodo and B.-S. Yang, ``Support vector machine in machine condition
  monitoring and fault diagnosis,'' \emph{Mechanical Systems and Signal
  Processing}, vol.~21, no.~6, pp. 2560--2574, 2007.

\bibitem{sparseAE}
Y.~{Qi}, C.~{Shen}, D.~{Wang}, J.~{Shi}, X.~{Jiang}, and Z.~{Zhu}, ``Stacked
  sparse autoencoder-based deep network for fault diagnosis of rotating
  machinery,'' \emph{IEEE Access}, vol.~5, pp. 15\,066--15\,079, 2017.

\bibitem{rzhao}
R.~Zhao, R.~Yan, Z.~Chen, K.~Mao, P.~Wang, and R.~X. Gao, ``Deep learning and
  its applications to machine health monitoring,'' \emph{Mechanical Systems and
  Signal Processing}, vol. 115, pp. 213 -- 237, 2019.

\bibitem{newfd1}
P.~Luo, Z.~Yin, D.~Yuan, F.~Gao, and J.~Liu, ``An intelligent method for early
  motor bearing fault diagnosis based on wasserstein distance generative
  adversarial networks meta learning,'' \emph{IEEE Transactions on
  Instrumentation and Measurement}, vol.~72, pp. 1--11, 2023.

\bibitem{newfd5}
H.~Wang, J.~Xu, R.~Yan, and R.~X. Gao, ``A new intelligent bearing fault
  diagnosis method using sdp representation and se-cnn,'' \emph{IEEE
  Transactions on Instrumentation and Measurement}, vol.~69, no.~5, pp.
  2377--2389, 2020.

\bibitem{EC_feature}
B.~Xue, M.~Zhang, W.~N. Browne, and X.~Yao, ``A survey on evolutionary
  computation approaches to feature selection,'' \emph{IEEE Transactions on
  Evolutionary Computation}, vol.~20, no.~4, pp. 606--626, 2016.

\bibitem{featureConst}
B.~Peng, Y.~Bi, B.~Xue, M.~Zhang, and S.~Wan, ``Multi-view feature construction
  using genetic programming for rolling bearing fault diagnosis [application
  notes],'' \emph{IEEE Computational Intelligence Magazine}, vol.~16, no.~3,
  pp. 79--94, 2021.

\bibitem{AutFeatureExtr}
B.~Peng, S.~Wan, Y.~Bi, B.~Xue, and M.~Zhang, ``Automatic feature extraction
  and construction using genetic programming for rotating machinery fault
  diagnosis,'' \emph{IEEE Transactions on Cybernetics}, vol.~51, no.~10, pp.
  4909--4923, 2021.

\bibitem{pratt}
L.~Y. Pratt, ``Discriminability-based transfer between neural networks,'' in
  \emph{Advances in Neural Information Processing Systems 5}, S.~J. Hanson,
  J.~D. Cowan, and C.~L. Giles, Eds.\hskip 1em plus 0.5em minus 0.4em\relax
  Morgan-Kaufmann, 1993, pp. 204--211.

\bibitem{sjPan}
S.~J. {Pan} and Q.~{Yang}, ``A survey on transfer learning,'' \emph{IEEE
  Transactions on Knowledge and Data Engineering}, vol.~22, no.~10, pp.
  1345--1359, Oct 2010.

\bibitem{sjPan2011}
S.~J. {Pan}, I.~W. {Tsang}, J.~T. {Kwok}, and Q.~{Yang}, ``Domain adaptation
  via transfer component analysis,'' \emph{IEEE Transactions on Neural
  Networks}, vol.~22, no.~2, pp. 199--210, Feb 2011.

\bibitem{longRTRL}
M.~Long, J.~Wang, G.~Ding, S.~J. Pan, and P.~S. Yu, ``Adaptation
  regularization: A general framework for transfer learning,'' \emph{IEEE
  Transactions on Knowledge and Data Engineering}, vol.~26, no.~5, pp.
  1076--1089, 2014.

\bibitem{lwen}
L.~{Wen}, L.~{Gao}, and X.~{Li}, ``A new deep transfer learning based on sparse
  auto-encoder for fault diagnosis,'' \emph{IEEE Transactions on Systems, Man,
  and Cybernetics: Systems}, vol.~49, no.~1, pp. 136--144, Jan 2019.

\bibitem{cross-domain}
X.~{Li}, W.~{Zhang}, and Q.~{Ding}, ``Cross-domain fault diagnosis of rolling
  element bearings using deep generative neural networks,'' \emph{IEEE
  Transactions on Industrial Electronics}, vol.~66, no.~7, pp. 5525--5534, July
  2019.

\bibitem{aks_quick}
A.~K. Sharma and N.~K. Verma, ``Quick learning mechanism with cross-domain
  adaptation for intelligent fault diagnosis,'' \emph{IEEE Transactions on
  Artificial Intelligence}, vol.~3, no.~3, pp. 381--390, 2022.

\bibitem{dagcn}
T.~Li, Z.~Zhao, C.~Sun, R.~Yan, and X.~Chen, ``Domain adversarial graph
  convolutional network for fault diagnosis under variable working
  conditions,'' \emph{IEEE Transactions on Instrumentation and Measurement},
  vol.~70, pp. 1--10, 2021.

\bibitem{NIPS2011}
J.~S. Bergstra, R.~Bardenet, Y.~Bengio, and B.~K\'{e}gl, ``Algorithms for
  hyper-parameter optimization,'' in \emph{Advances in Neural Information
  Processing Systems 24}, J.~Shawe-Taylor, R.~S. Zemel, P.~L. Bartlett,
  F.~Pereira, and K.~Q. Weinberger, Eds.\hskip 1em plus 0.5em minus 0.4em\relax
  Curran Associates, Inc., 2011, pp. 2546--2554.

\bibitem{BJ}
J.~Bergstra and Y.~Bengio, ``Random search for hyper-parameter optimization,''
  \emph{J. Mach. Learn. Res.}, vol.~13, pp. 281--305, Feb. 2012.

\bibitem{BB_nn_rein}
B.~Baker, O.~Gupta, N.~Naik, and R.~Raskar, ``Designing neural network
  architectures using reinforcement learning,'' \emph{CoRR}, vol.
  abs/1611.02167, 2016.

\bibitem{c1rv1}
\BIBentryALTinterwordspacing
J.~Fang, Y.~Sun, Q.~Zhang, Y.~Li, W.~Liu, and X.~Wang, ``Densely connected
  search space for more flexible neural architecture search,'' in \emph{2020
  IEEE/CVF Conference on Computer Vision and Pattern Recognition (CVPR)}.\hskip
  1em plus 0.5em minus 0.4em\relax Los Alamitos, CA, USA: IEEE Computer
  Society, jun 2020, pp. 10\,625--10\,634. [Online]. Available:
  \url{https://doi.ieeecomputersociety.org/10.1109/CVPR42600.2020.01064}
\BIBentrySTDinterwordspacing

\bibitem{sun_pso}
Y.~{Sun}, B.~{Xue}, M.~{Zhang}, and G.~G. {Yen}, ``A particle swarm
  optimization-based flexible convolutional autoencoder for image
  classification,'' \emph{IEEE Transactions on Neural Networks and Learning
  Systems}, vol.~30, no.~8, pp. 2295--2309, Aug 2019.

\bibitem{Ysun2020}
Y.~Sun, B.~Xue, M.~Zhang, and G.~G. Yen, ``Evolving deep convolutional neural
  networks for image classification,'' \emph{IEEE Transactions on Evolutionary
  Computation}, vol.~24, no.~2, pp. 394--407, 2020.

\bibitem{EvoDCNN}
\BIBentryALTinterwordspacing
T.~Hassanzadeh, D.~Essam, and R.~Sarker, ``Evodcnn: An evolutionary deep
  convolutional neural network for image classification,''
  \emph{Neurocomputing}, vol. 488, pp. 271--283, 2022. [Online]. Available:
  \url{https://www.sciencedirect.com/science/article/pii/S092523122200145X}
\BIBentrySTDinterwordspacing

\bibitem{PHAN2024101573}
Q.~M. Phan and N.~H. Luong, ``Parameter-less pareto local search for
  multi-objective neural architecture search with the interleaved multi-start
  scheme,'' \emph{Swarm and Evolutionary Computation}, vol.~87, p. 101573,
  2024.

\bibitem{Ian}
T.~Chen, I.~J. Goodfellow, and J.~Shlens, ``Net2net: Accelerating learning via
  knowledge transfer,'' \emph{CoRR}, vol. abs/1511.05641, 2015.

\bibitem{tl-gdbn}
G.~{Wang}, J.~{Qiao}, J.~{Bi}, W.~{Li}, and M.~{Zhou}, ``Tl-gdbn: Growing deep
  belief network with transfer learning,'' \emph{IEEE Transactions on
  Automation Science and Engineering}, vol.~16, no.~2, pp. 874--885, April
  2019.

\bibitem{cwru}
W.~A. Smith and R.~B. Randall, ``Rolling element bearing diagnostics using the
  {Case Western Reserve University} data: A benchmark study,'' \emph{Mechanical
  Systems and Signal Processing}, vol.~64, pp. 100--131, 2015.

\bibitem{paderborn}
C.~Lessmeier, J.~Kimotho, D.~Zimmer, and W.~Sextro,
  ``\BIBforeignlanguage{en}{Condition monitoring of bearing damage in
  electromechanical drive systems by using motor current signals of electric
  motors: A benchmark data set for data-driven classification},''
  \emph{\BIBforeignlanguage{en}{European Conf., PHM Society, Bilbao (Spain)}},
  vol.~3, no.~1, 2016.

\bibitem{gfd}
\BIBentryALTinterwordspacing
Y.~Pandya, ``Gearbox fault diagnosis data,'' 06 2018. [Online]. Available:
  \url{https://data.openei.org/submissions/623}
\BIBentrySTDinterwordspacing

\bibitem{moea2}
N.~Srinivas and K.~Deb, ``Muiltiobjective optimization using nondominated
  sorting in genetic algorithms,'' \emph{Evol. Comput.}, vol.~2, no.~3, p.
  221–248, 1995.

\bibitem{nsga}
K.~Deb, A.~Pratap, S.~Agarwal, and T.~Meyarivan, ``A fast and elitist
  multiobjective genetic algorithm: {NSGA-II},'' \emph{IEEE Transactions on
  Evolutionary Computation}, vol.~6, no.~2, pp. 182--197, 2002.

\bibitem{nsga-iii}
K.~Deb and H.~Jain, ``An evolutionary many-objective optimization algorithm
  using reference-point-based nondominated sorting approach, part {I}: Solving
  problems with box constraints,'' \emph{IEEE Transactions on Evolutionary
  Computation}, vol.~18, no.~4, pp. 577--601, 2014.

\bibitem{10807400}
Y.~Li, Z.~Xie, S.~Yang, and Z.~Ren, ``A novel hybrid multi-objective
  optimization algorithm and its application to designs of electromagnetic
  devices,'' \emph{IEEE Transactions on Magnetics}, vol.~61, no.~2, pp. 1--4,
  2025.

\bibitem{rkhs}
N.~Aronszajn, ``Theory of reproducing kernels,'' \emph{Transactions of the
  American Mathematical Society}, vol.~68, no.~3, pp. 337--404, 1950.

\bibitem{bengio}
Y.~Bengio, P.~Lamblin, D.~Popovici, and H.~Larochelle, ``Greedy layer-wise
  training of deep networks,'' in \emph{Advances in Neural Information
  Processing Systems 19}, B.~Sch\"{o}lkopf, J.~C. Platt, and T.~Hoffman,
  Eds.\hskip 1em plus 0.5em minus 0.4em\relax MIT Press, 2007, pp. 153--160.

\bibitem{friedmann1940}
M.~Friedman, ``A comparison of alternative tests of significance for the
  problem of m rankings,'' \emph{The Annals of Mathematical Statistics},
  vol.~11, no.~1, pp. 86--92, 1940.

\bibitem{sheskin2003handbook}
D.~J. Sheskin, \emph{Handbook of {Parametric} and {Nonparametric} {Statistical}
  {Procedures}: {Third} {Edition} (3rd ed.)}.\hskip 1em plus 0.5em minus
  0.4em\relax Chapman and Hall/CRC, 2003.

\bibitem{bdtest}
O.~J. Dunn, ``Multiple comparisons among means,'' \emph{Journal of the American
  Statistical Association}, vol.~56, no. 293, pp. 52--64, 1961.

\bibitem{stat_cd}
J.~Dem\v{s}ar, ``Statistical comparisons of classifiers over multiple data
  sets,'' \emph{J. Mach. Learn. Res.}, vol.~7, p. 1–30, 12 2006.

\end{thebibliography}

\begin{IEEEbiography}[{\includegraphics[width=1in,height=1.5in,clip,keepaspectratio]{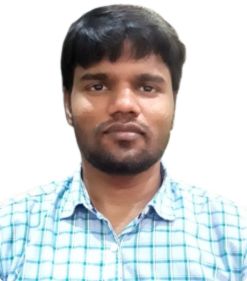}}]{Arun K. Sharma} received his PhD in Electrical Engineering from Indian Institute of Technology, Kanpur in 2024. Dr. Arun received his B.Tech. in Electrical Engineering from BIT Sindri Dhanbad and M.Tech. in Instrumentation from Indian Institute of Technology, Kharagpur. He is currently working as a post-doctoral researcher at SMSS Lab, Mechanical Engineering, IIT Kanpur.  His research interests include signal processing, deep learning for pattern recognition and analysis, sustainable AI, and AI in advanced engineering applications. He has been a reviewer for several reputed journals, such as the IEEE Transactions on Aerospace and Electronic Systems (TAES), the IEEE Transactions on Fuzzy Systems, and the IEEE Computational Intelligence Magazine.
\end{IEEEbiography}

\begin{IEEEbiography}[{\includegraphics[width=1in,height=1.5in,clip,keepaspectratio]{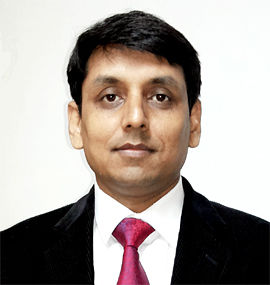}}]{Nishchal K. Verma} (Senior Member, IEEE) received the Ph.D. degree in electrical engineering from the Indian Institute of Technology Delhi, India, in 2007. He is currently a Professor with Electrical Engineering, IIT Kanpur, India. He has published over 270 research papers in reputable journals and conferences and has authored or co-authored six books. He has successfully completed research projects from various funding agencies such as The BOEING Company (USA), DRDO, JCBCAT, MHRD, CSIR, IIT Kanpur, MCIT, SFTIG, and VTOL. His research interests include AI-related theories and its practical applications to interdisciplinary domains. Dr. Verma is an awardee of Devendra Shukla Young Faculty Research Fellowship (2013–2016) by the IIT Kanpur, India. Dr. Verma serves as an Associate Editor for key publications, including IEEE Transactions on Artificial Intelligence (2020-2024), IEEE Transactions on Neural Networks and Learning Systems (2019-2024), and IEEE Computational Intelligence Magazine (2015-2020).
\end{IEEEbiography}

\end{document}